\documentclass[12pt,conference,letterpaper,onecolumn]{IEEEtran}
\pdfoutput=1
\usepackage{times}
\usepackage{amsmath}
\usepackage{amsfonts}
\usepackage{amssymb}
\usepackage{xcolor}
\usepackage{algorithm}
\usepackage{algpseudocode}
\usepackage{ifdraft}

\usepackage[pdftex]{graphicx}
\graphicspath{./figures/}
\DeclareGraphicsExtensions{.pdf}

\usepackage{float}
\usepackage{subfig}

\ifoptionfinal{
\newcommand{\note}[1]{}
\newcommand{\ysnote}[1]{}
\newcommand{\aknote}[1]{}
\newcommand{\cwnote}[1]{}
\newcommand{\mfnote}[1]{}
} {
\newcommand{\ysnote}[1]{  {\textcolor{magenta}      { ***Yogesh: #1 }}}
\newcommand{\aknote}[1]{  {\textcolor{blue}      { ***Alok: #1 }}}
\newcommand{\cwnote}[1]{  {\textcolor{brown}      { ***Charith: #1 }}}
\newcommand{\mfnote}[1]{  {\textcolor{teal}      { ***Marc: #1 }}}
\newcommand{\note}[1]{  {\textcolor{red}      { ***#1 }}}
}

\title{Scalable Analytics over Distributed Time-series Graphs using \emph{GoFFish}}

\author{%
{Yogesh Simmhan$^{\dagger}$, Charith Wickramaarachchi, Alok Kumbhare, Marc Frincu, }\\
{Soonil Nagarkar, Santosh Ravi, Cauligi Raghavendra, Viktor Prasanna}%

\vspace{1.6mm}\\
\fontsize{10}{10}\selectfont\itshape
$^{\dagger}$Indian Institute of Science, Bangalore 560012 India\\
University of Southern California, Los Angeles CA 90089 USA\\
\fontsize{9}{9}\selectfont\ttfamily\upshape

\vspace{1.2mm}\\
simmhan@serc.iisc.in, \{cwickram,kumbhare,frincu,snagarka,sathyavi,raghu,prasanna\}@usc.edu
\fontsize{10}{10}\selectfont\rmfamily\itshape
}

\begin{document}
\maketitle
\begin{abstract} 
Graphs are a key form of Big Data, and performing scalable analytics over them is invaluable to many
domains. As our ability to collect data grows, there is an emerging class of inter-connected data
which accumulates or varies over time, and on which novel analytics -- both over the network
structure and across the time-variant attribute values -- is necessary. We introduce the notion of
\emph{time-series graph analytics} and propose \emph{Gopher}, a scalable programming abstraction to
develop algorithms and analytics on such datasets. Our abstraction leverages a sub-graph centric
programming model and extends it to the temporal dimension using an \emph{iterative BSP} (Bulk Synchronous
Parallel) approach. Gopher is co-designed with \emph{GoFS}, a distributed storage specialized for
time-series graphs, as part of the \emph{GoFFish} distributed analytics platform. We examine storage
optimizations for GoFS, design patterns in Gopher to leverage the distributed data layout, and
evaluate the GoFFish platform using time-series graph data and applications on a commodity cluster.
\end{abstract}

\section{Introduction}
\label{sec:intro}
With the proliferation of ubiquitous physical devices (e.g. urban monitoring, smart power meters) and
virtual agents (e.g. Twitter feeds, Foursquare check-ins) that sense, monitor and track
human and environmental activity, data is streaming more continuously and is intrinsically
interconnected. Two defining characteristics of such datasets, endemic to both the Internet of Things~\cite{iot}
and Social Networks, are the temporal or time-series attributes and the topological relationships that
exist between them. Such datasets that imbue both these temporal and graph
features have not been adequately examined from the perspective of scalable Big Data management and
analysis, even as they are becoming pervasive.

\begin{figure}
\centering
\includegraphics[width=0.7\columnwidth]{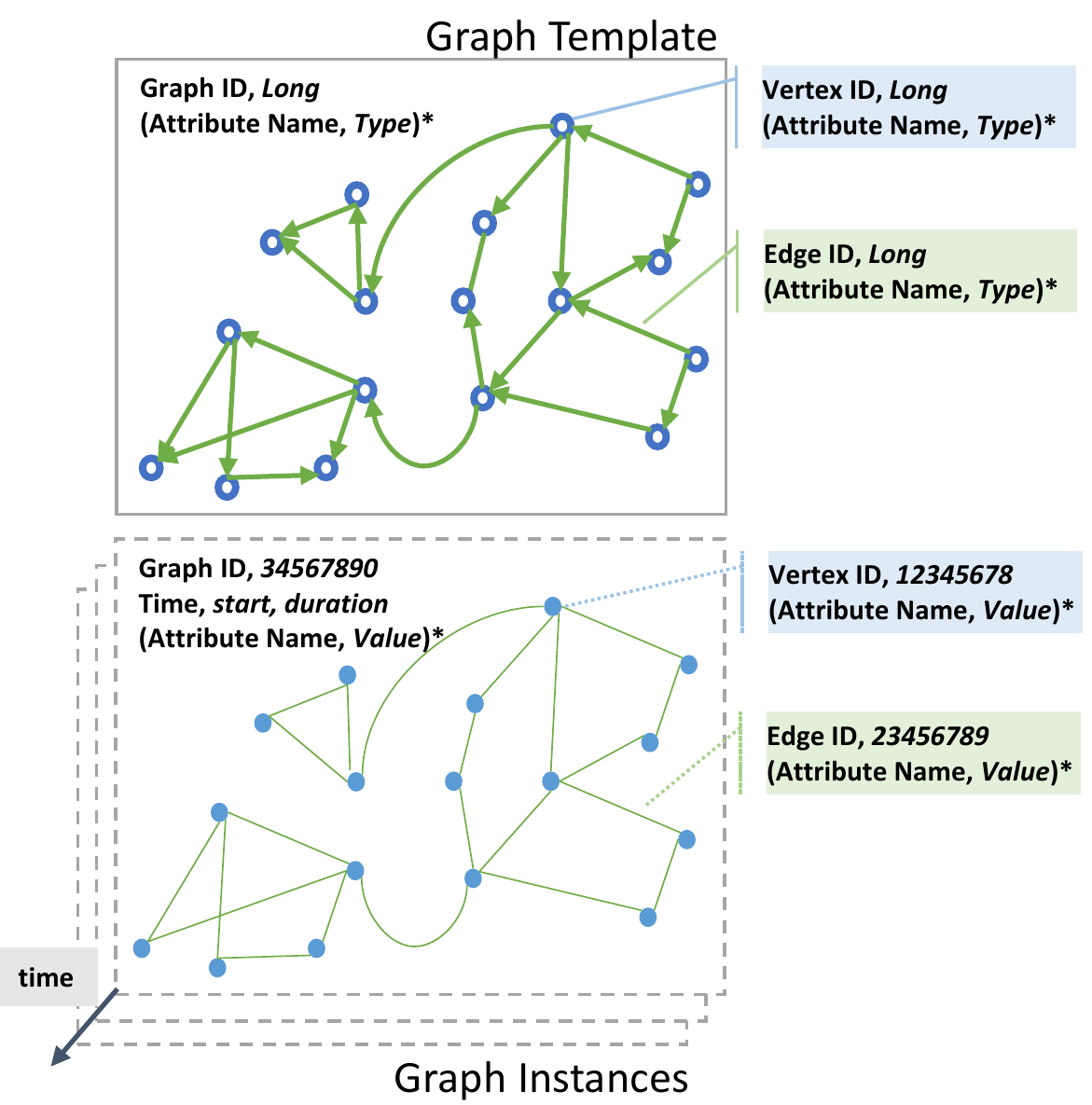}
\caption{Time-series graph collection. \emph{Template} captures static/slow-changing topology and attribute names. \emph{Instances} record
  temporally variant attribute values for this template.}
\label{fig:tsgraphs}
\vspace{-0.2in}
\end{figure}

Graph datasets with temporal characteristics have been variously known in literature as temporal
graphs~\cite{kostakos:temporal}, kineographs~\cite{cheng:eurosys:2012}, dynamic
graphs~\cite{cortes:dynamic:2004} and time-evolving graphs~\cite{tong:sdm:2008}. 
Temporal graphs capture the time variant network structure in a single graph by introducing a
temporal edge between the same vertex at different moments. Others construct graph snapshots at
specific change points in the graph structure. In particular, kineographs deal with graph that
exhibit high dynamism.
In this paper, we focus on a related class of \emph{time-series graphs} which we define to be graphs whose network topology is
slow-changing but whose attribute values associated with vertices and edges change (or are generated) more
frequently. As a result, we have a series of graphs, each of whose vertex and edge attributes capture the
instantaneous state of a system at a point in time (e.g. 2013-07-14
15:00~PST), or the cumulative states of the system over
time durations (e.g. 2013-07-14 15:00~PST --
15:05~PST), but whose numbers of, and connectivity between, vertices and edges are less
dynamic. Each graph in the time-series is an \emph{instance}, while the slow changing topology is a \emph{template} (Fig.~\ref{fig:tsgraphs}).

Many emerging ``Big Data'' datasets can be captured using such a time-series graph model. For e.g., when we
consider images captured by a network of traffic or surveillance cameras in a large
city like Los Angeles or London, these cameras themselves are inter-connected by the roads that link
them (i.e. the graph topology) and their location changes less often. 
On the other hand, the periodic data that they generate after image processing, such
as \emph{vehicle license IDs} at the vertices and the current \emph{travel-time} on edges, form a time-series
graph. 
Similarly, we can construct time-series graphs
out of social network feeds (friend network forms the ``slow changing'' topology while tweets or
posts form the instances) or even Internet trace-route statistics
(Internet IPs on network hops form the topology while current latency and bandwidth form
instances). As such, these datasets are expected to have over $O(10^6)$ vertices, $O(10^7)$
edges, and $O(10^4)$ time points.

Many innovative analytics applications can leverage the topological and temporal information that a time-series
graph representation provides.  While some are a temporal extension of existing algorithms, others
are more novel. For e.g., we can extend the Dikstra's shortest path
algorithm to a temporal version over a road network with snapshots of historical
traffic conditions accumulated every 5-min.  We start traversing at the source vertex of a graph
instance and use its travel times in estimating the shortest path. But after traveling
5-mins and reaching the \emph{temporal boundary} of that graph instance (e.g. we determine that the next position cannot be in the current graph instance due to traffic conditions such as the average speed), we switch over to the next
graph instance in the time-series with traffic information on the next 5-mins, and resume
traversal. This gives us concentric waves of traversals. 
The space of such
large-scale temporal graph analytics applications is rich, ranging from studying time-evolving social network
communities to Internet router bottlenecks.

In our recent work~\cite{simmhan:ipdps:2014}, we extended the idea of scalable vertex centric Bulk
Synchronous Parallel (BSP)~\cite{valiant:bsp} graph programming model, championed by Google's
Pregel~\cite{pregel} and implemented by others like Giraph~\cite{giraph,gps}, into a \emph{sub-graph centric} model, \emph{Gopher}, that was targeted at
commodity cluster and Cloud platforms. The platform operated on a single graph and significantly
out-performed Giraph in our benchmarks. In this paper, we further expand Gopher to
support time-series graph analysis in distributed environments. Specifically, we propose structured
programming abstractions for large-scale analytics across multiple graphs using an \emph{iterative BSP} model.  In addition, we
couple this with a distributed storage model optimized for time-series graphs called
\emph{Graph-oriented File System (GoFS)}, that we introduce here. GoFS partitions the graph based on
topology, groups instances over time on disk, and allows attribute-value pairs to be associated
with the vertices and edges of the instances. In particular, GoFS is designed to scale out for
common data access patterns over time-series graphs, and this data layout is intelligently leveraged
by Gopher during execution. Gopher and GoFS are part of the GoFFish graph analytics framework.

We make the following specific contributions in this paper:
\begin{enumerate}
\item We propose a time-series graph model and an iterative Bulk Synchronous Parallel
  programming abstraction for composing sub-graph centric analytics over this data model (\S~\ref{sec:temporal}--\ref{sec:programming});
\item We discuss design patterns for classes of time-series graph analytics, and
  present a distributed graph storage layout optimized for these access patterns (\S~\ref{sec:gofs});
\item We implement the iterative BSP abstraction and distributed layout as part of the GoFFish
  analytics platform, and present an empirical evaluation of the abstractions and optimizations for sample
  graph algorithms and datasets (\S~\ref{sec:perf}). 
\end{enumerate}


\section{Related Work}

With the advent of truly Big data sources with graph oriented structures in modern computing, there
has been a renewed focus on scalable and accessible platforms for their analysis. Current approaches
can be separated into three categories; Map/Reduce frameworks, Bulk Synchronous Parallel (BSP)
frameworks, and online processing systems. Our effort is focused on BSP and for offline
queries. While many of the mature frameworks offer complete distributed solutions, with fault
tolerance and performance guarantees, GoFFish focuses on exploring novel graph processing
abstractions that are easy to use and can be scaled. In this paper, we do not reinvent well
understood reliability and recovery approaches.

Map/Reduce~\cite{dean2008mapreduce} has become a \emph{de facto} abstraction for large data
analysis, and has been applied to graph data as well~\cite{Kang:2011:HMR}. However its use for
general purpose graph processing can lead to both performance and usability
concerns~\cite{cohen2009cise}.  Iterative versions of Map/Reduce have been proposed to address some
of the limitations of the original abstraction~\cite{twister}. Further extensions have
added features for aggregation~\cite{pike2005interpreting} and SQL-like queries, but these are ill suited
for graph processing, which is better served through a message passing
model~\cite{lumsdaine2007challenges}. However, it is possible to improve upon these frameworks
through graph-specific programming models. GoFFish is in this vein, combining
prior work on BSP graph frameworks with timeseries analysis using a sub-graph centric
model~\cite{simmhan:ipdps:2014,redekopp2011performance}. 

Google recently proposed the Pregel model~\cite{pregel} of vertex centric programming for large
scale graph processing, which also has similarities with GraphLab~\cite{low2012distributed}. The Pregel model allows the programmer to implement algorithms from the point of view of a single vertex which
remarkably simplifies the programming model for a large class of graph algorithms and allows for much simpler and quicker development. Further, Pregel's execution model is based on BSP~\cite{valiant:bsp}, where computation
is done through a series of barriered iterations called supersteps. This model of parallel computing
eliminates concerns of deadlocks and data races common in asynchronous
systems~\cite{lumsdaine2007challenges}. While the original BSP model fell out of favor in parallel computing due to the large cost of superstep synchronization, for large graphs, one can balance machine-level vertex computation load and use hierarchical synchronization to mitigate this synchronization overhead. Further, with the push toward commodity hardware and Cloud infrastructure, the emphasis is more on scalability than performance on HPC hardware.

The vertex-centric BSP model has been adopted by a variety of
frameworks~\cite{redekopp2011performance} like Giraph~\cite{giraph}, Hama and GPS~\cite{gps} that
implement improvements on the original idea. Giraph is currently the popular framework in this
space, being adopted by Facebook for large scale graph analysis on their network data. GPS is a
vertex-centric framework with support for dynamic repartitioning of the graph between hosts. Many of
these frameworks offer engineering optimizations to improve performance and simplify programming.
These include master compute methods, send- and receive-side message aggregation, dynamic partition
balancing, and message as well as graph memory compression techniques. Many of these optimizations focus on
reducing the effective number of messages passed around the system, both in memory and on the
network. This emphasis is because the number of messages correspond roughly to the number of edges
for a large class of graph algorithms. On large graphs with power law out-degree distributions, the
number of generated messages can flood both memory and network. However, part of the problem lies
not in engineering solutions but in that these frameworks do not deviate much from Pregel's original
vertex centric model. This limits more significant optimizations possible within them.

Our earlier work on GoFFish improves on Pregel by proposing a subgraph centric model rather than
a vertex centric one~\cite{simmhan:ipdps:2014}. For a large number of algorithms the amount of
work performed per vertex is so negligible that the overhead of massive parallelism can outweigh the benefits. By using a subgraph as a unit of computation, we show that the efficiency of
every worker is increased, and the number of messages the framework must handle is dramatically reduced,
since it is more a function of the number of unique edges between sub-graphs that span partitions, rather than between
vertices. This also results in effectively more work being performed in each
superstep, and thus requires fewer supersteps, with associated synchronization overhead, to complete the application.

This comes at the cost of marginally increasing the complexity of the programming model, mixing
features of vertex centric and shared-memory graph abstractions. But for many applications the
performance and scalability improvements may be worth the costs. In this article, we further expand
upon this to support time-series graphs, which Pregel does not natively support and is punitive to
implement na\"{i}vely using the vertex centric approach. We also investigate a novel distributed
data storage that is optimized for time-series graphs. While Pregel does not prescribe any data
storage, the Apache Giraph implementation of Pregel retains the tuple-based HDFS for storing graphs,
which impacts initial loading from disk to memory even for single graphs.

Online processing systems such as Kineograph~\cite{cheng:eurosys:2012} and
Trinity~\cite{shao2013trinity} are graph processing models that focus heavily on the analysis of
streaming information, and are thus purpose built for time evolving graphs. These systems are able
to process an large quantity of information with timeliness guarantees. Kineograph's approach can
also potentially support time-series graphs using consistent snapshots with an epoch commit
protocol. Traditional graph algorithms are then run on top of each static snapshot. However, GoFFish
does not aim to provide streaming or online graph processing services, but rather more traditional
offline bulk processing on large datasets. Dealing with dynamic topologies and streaming data is not
within the scope of this paper.





\section{Analytics over Time-series Graphs}
\label{sec:temporal}

\subsection{Time-series Graphs}
Time-series graphs can be considered as snapshots of a graph recorded over time
(Fig.~\ref{fig:tsgraphs}).  We define a \emph{collection} of time-series graphs as $\Gamma = \langle
\widehat{G}, G \rangle$, where $\widehat{G}$ is called a \emph{graph template} that is the time
invariant topology, and $G$ is an ordered set of \emph{graph instances}, capturing time-variant
values. $\widehat{G} = ( \widehat{V}, \widehat{E} )$ gives the set of vertices, $\hat{v}_i \in
\widehat{V}$, and edges, $\hat{e}_j \in \widehat{E}:\widehat{V}\rightarrow\widehat{V}$, common to
all instances. The graph
instance $g^t \in G$ at timestamp $t$ is given by $( V^t, E^t, t )$ where $V$ and $E$ capture the vertex
and edge \emph{values} for $\widehat{V}$ and $\widehat{E}$ at time $t$, $|V^t|=|\widehat{V}|$ and $|E^t|=|\widehat{E}|$. The set $G$ is ordered in time.

Vertices and edges in the template have a defined set of typed attributes, 
$\mathbb{A}(\widehat{V})=\{id, \alpha_1, \ldots, \alpha_m\}$ and $\mathbb{A}(\widehat{E})=\{id, \beta_1, \ldots, \beta_n\}$
respectively. All vertices and edges share the same set of attributes with $id$ being one of the unique identifier attribute.
These similar attributes are also present in the instances, except for the $id$
attribute, which is set in the template
Thus each vertex $v^{t}_{i} \in V^t$
for a graph instance $g^t$ at time $t$ has a set of attribute values $\{\hat{v}_{i}.id, v_{i}^{t}.\alpha_{k},
\ldots, v_{i}^{t}.\alpha_{m}\}$, and each edge $e^{t}_{j} \in E^t$ has attribute values
$\{\hat{e}_{j}.id, e_{j}^{t}.\beta_{l}, \ldots, e_{j}^{t}.\beta_{n}\}$,

A slow changing graph topology can be captured using the special \emph{isExists} attribute flag that allows us to simulate the appearance/disappearance of vertices or edges throughout the time-series. 

\subsection{Sample Applications}

Analytics over individual graphs fall broadly into traversal (e.g. shortest path under changing conditions), centrality
detection (e.g. betweenness centrality at different points in time) and clustering algorithms (e.g. evolution of community), among
others. Applications over time-series graphs expand on these possibilities.

\textbf{Centrality detection} algorithms form a class of applications where the state of a vertex is
analyzed in raport with the rest. Example applications include the page rank algorithm where each
vertex's rank relative to the other vertices' rank is analyzed at each time snapshot.  Since each
graph instance contains all the information required to determine if a vertex is present in a path,
at \emph{a particular moment}, such analytics can operate on each instance independently. In that
sense, they are similar to algorithms for individual graphs, except that they are repeated for each
graph instance.

\textbf{Clustering} algorithms discover the existence of localized, time-evolving
patterns. While each pattern can be identified independently for each instance, the individual
results need to be aggregated at the end of the execution to get the global view. Applications that
can pe placed in this category range from studies on the PageRank stability over time to analyzing
the dynamics of a person's social network and identifying frequent clusters in gene expression
networks. Here, the application initially operates on each instance independently, but has to
eventually perform an aggregation or analysis that spans the synthesized result from each instance. 

\textbf{Traversal} algorithms show a linear dependence between instances as information
gathered in the past drives traversals in the future. The shortest path over time-varying traffic
conditions, presented in \S~\ref{sec:intro}, falls in this space, and can be applied to network
packet tracing or meme propagation. Further, this class includes minimum spanning trees to determine
the optimal route for patrolling, and epidemiological studies to find the time for a disease to
spread. Here, there is either a strict sequential dependence between one instance and its
predecessor, or using information from a prior instance will help efficiently localize the search on
the next.


\subsection{Design Patterns for Graph Analytics over Time}
\begin{figure}
\centering
\includegraphics[width=\columnwidth]{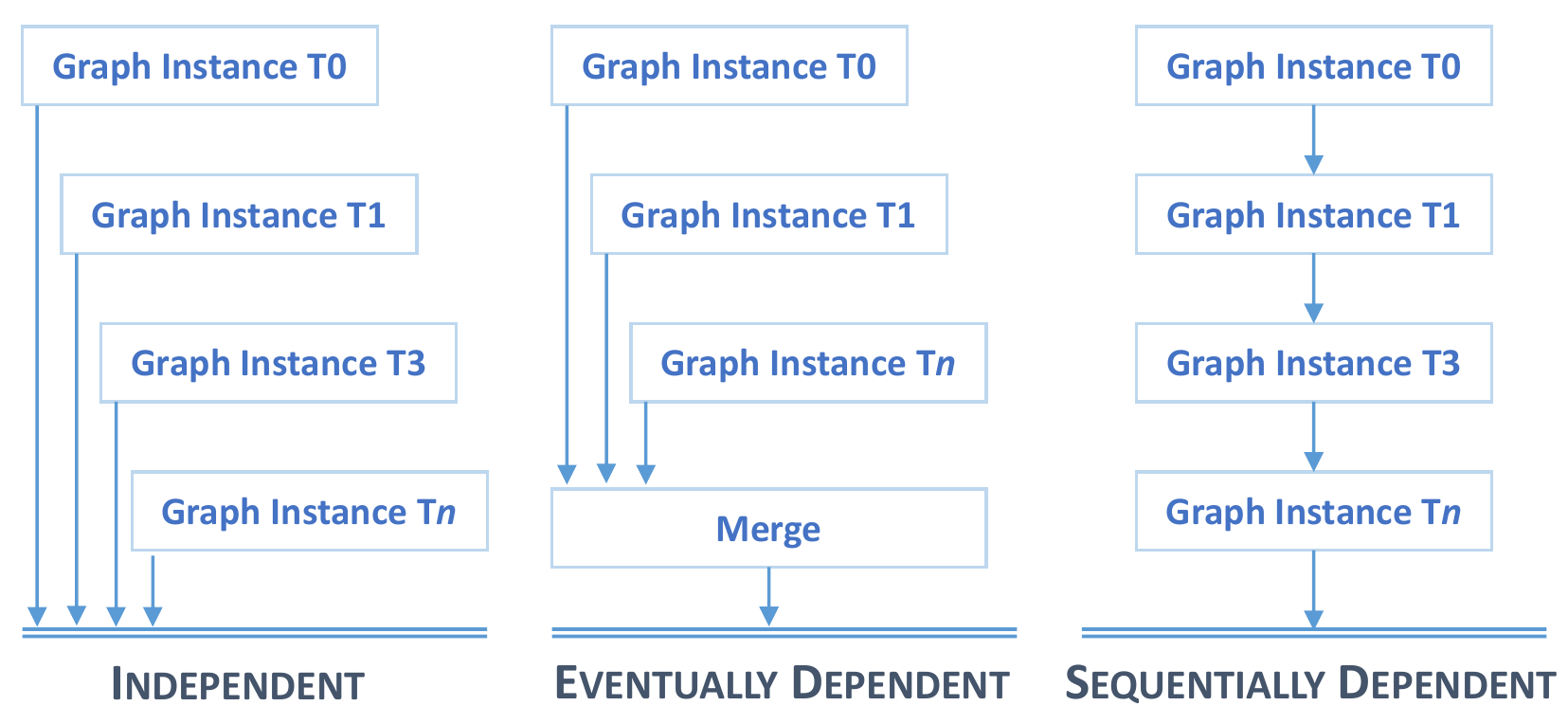}
\caption{Design patterns for analytics on time-series graphs.}
\label{fig:models}
\vspace{-0.2in}
\end{figure}

Based on these sample classes of algorithm we can synthesize three types of composition patterns for
temporal graph analytics. These are illustrated in Fig.~\ref{fig:models} and described next.
\begin{enumerate}
\item Analysis over every graph instance is \textbf{independent}. The result from the application is just a
  union of results from each graph instance;
\item Graph instances are \textbf{eventually dependent}. Each instance can execute independently but results from
  all instances need to be aggregated or summarized to produce the final result; 
\item Graphs instances are \textbf{sequentially dependent}. Here, analysis over a graph instance
  cannot start (or, as a variation, complete) before the results from the previous graph instance
  are available. 
\end{enumerate}

These patterns have two purposes in mind: make it \emph{easy} for users to design common time-series
graph analytics, and make it possible to \emph{efficiently scale} them in a distributed environment.
In the abstraction
(\S~\ref{sec:programming}) and evaluation sections (\S~\ref{sec:perf}) we show exemplar applications
mapped to these three patterns and empirically analyze them. 

The independent pattern is similar to a \texttt{Parallel For-Each} construct. It allows concurrent
execution over each graph instance with the parallelism that can be ideally exploited being equal
to the number of graph instances. 

The eventually dependent pattern captures the \texttt{Fork-Join} paradigm. This brings in additional
synchronization capability for aggregation without compromising the parallelism. Pipelining can be facilitated if the pattern is
extended to an ``incremental'' join where the \emph{Merge} step starts as soon as the first graph
instance completes, and ends when the last instance has finished executing. 

The sequentially dependent pattern is a traditional sequential execution model. While it does not
offer concurrency across time (i.e. instances), as we shall see, the subgraph centric BSP
abstraction allows for spatial concurrency across vertices within a graph instance.

These models are not meant to be a comprehensive list and serve as the building blocks to help construct a large class of applications while
retaining concurrency. 
While they can be incrementally extended to more complex patterns, e.g., monotonic
dependency (a relaxation of sequential dependency), random access across instances, etc. we
omit discussing these cases in this paper in the interests of being concise.



\section{Programming Time-series Graph Analytics}
\label{sec:programming}

In developing abstractions to fit the design patterns from \S~\ref{sec:temporal}, we build upon our
recent work on \emph{sub-graph centric programming abstractions} targeted at distributed programming over
single graphs. We first present this initial model and then extend it with our
novel \emph{iterative BSP} model, along with several time-series graph algorithms mapped to it.

\subsection{Sub-graph Centric Programming Abstraction}
A sub-graph centric distributed programming abstraction defines the graph application logic from the
perspective of a single sub-graph within a partitioned graph. A graph $G=(V, E)$ is partitioned into
$n$ partitions, $\langle P_1=(V_1, E_1), \cdots, P_n=(V_n, E_n) \rangle$ such that
$\bigcup\limits_{i=1}^{n}V_i = V$, $\bigcup\limits_{i=1}^{n}E_i = E$, and $\forall i \ne j:V_i \cap
V_j = \varnothing$, i.e. a vertex is present in only one partition, and all edges appear in 
one partitions, with the exception being ``remote'' edges that can span
two partitions; $R_i=\{e_p | e_p \in E_i \text{ and } \forall e_p:v_q \rightarrow v_r, v_q \in V_i
\text{ and } v_r \not\in V_i \}$, if the edges are directed. If undirected, then either the source
or the sink vertex is present is another partition. Conversely, ``local'' edges for a partition are
those edges that are not remote; $L_i=\{e_p | e_p \in E_i \text{ and } \forall e_p:v_q \rightarrow
v_r, \text{ both } v_q,v_r
\in V_i \}$. Typically, partitioning tries to ensure that the number of
vertices, $|V_i|$, is equal across partitions and the total number of remote edges,
$\sum_{i=1}^{n}|R_i|$, is minimized.

Given this, a \emph{\textbf{sub-graph}} within a partition is a maximal set of vertices that are connected
through ``local'' edges. A partition $i$ may have between one and $|V_i|$ sub-graphs.

\begin{figure}
\centering
\includegraphics[width=0.8\columnwidth]{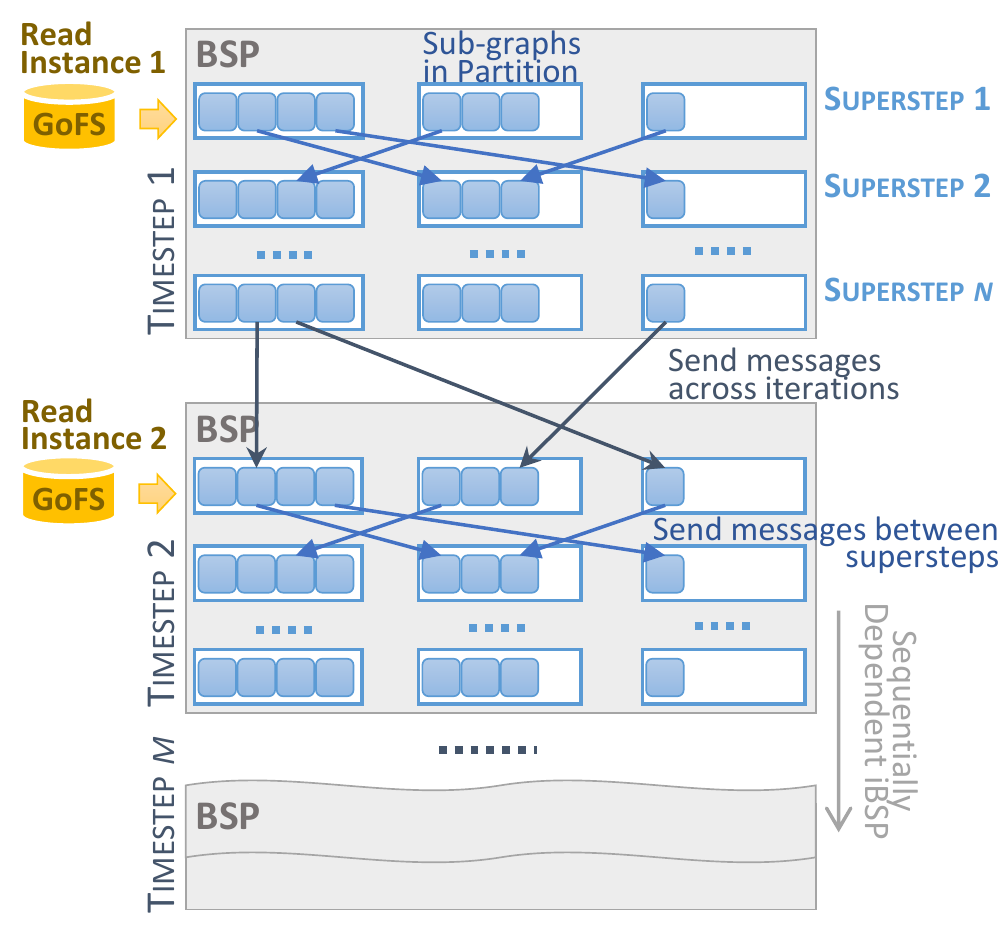}
\caption{Iterative BSP programming model across time-series graph instances. Each BSP timestep
  operates on a single graph instance, and is itself decomposed into multiple
  supersteps as part of the sub-graph centric abstraction. Graph instance data access takes place at the start of
  each BSP timestep.}
\label{fig:iBSP}
\vspace{-0.2in}
\end{figure}

In sub-graph centric programming, the user defines an application logic as a \texttt{Compute} method
that operates on a single sub-graph, independently. The method, upon completion, can exchange
messages with other sub-graphs, typically those that share a remote edge with the source sub-graph. A
single execution of the \texttt{Compute} method on all sub-graphs, each of which can execute
concurrently, forms a \emph{superstep}. Execution proceeds as a series of coordinated supersteps,
executed in a Bulk Synchronous Parallel (BSP) model. Messages generated in one superstep are
transmitted in ``bulk'' between supersteps, and available to the \texttt{Compute} of the destination
sub-graph in the next superstep. Execution stops when all \texttt{Compute} methods \emph{Vote to
Halt} and there are no messages generated within a superstep. Fig.~\ref{fig:iBSP} illustrates this
execution model.

The sub-graph centric programming abstraction is itself an extension to the vertex centric model,
where the \texttt{Compute} logic is from the perspective of a single vertex~\cite{pregel}. Our prior work~\cite{simmhan:ipdps:2014} shows
the performance benefits of this innovation by limiting both the number of supersteps requiring costly
synchronization and the number of messages exchanged. It also ease the programmability 
through the reuse of efficient, shared-memory graph algorithms within a single sub-graph.

\subsection{Iterative BSP for Time-series Graph Programming}
Sub-graph centric BSP programming offers natural parallelism across the graph topology. But it
operates on a single graph instance. In a sense, a single BSP execution corresponds to one box in
Fig.~\ref{fig:models} that operates on a single graph instance. Here, we use BSP as a building block
to define an \emph{iterative BSP (iBSP)} abstraction that meets the design patterns proposed
before. An iBSP application is a set of BSP steps, also referred to as \emph{timesteps} since each
operates on a single graph instance in time. While the BSP timestep itself can be opaque, we use the
sub-graph centric abstraction consisting of supersteps as its constituent. In a sense, the timestep
iteration acts as an ``outer loop'' while the supersteps over sub-graphs represent the ``inner loop''. The
execution order of the timesteps and the messages passed between them decides the iBSP
application's design pattern as in \S~\ref{sec:temporal}.

\textbf{Orchestration and Concurrency.} An iBSP application operates over a graph collection, $\Gamma$, which, as we defined earlier, is a
list of time ordered graph instances. 
As before, the users implement a \texttt{Compute} method which is invoked on every sub-graph \emph{and for every graph instance}. In case of an eventually
dependent pattern, an optional \texttt{Merge} method is available for invocation after all
instance timesteps complete.

For a \emph{sequentially dependent} pattern, only one graph instance and hence BSP timestep is
active at a time. The \texttt{Compute} method is called on all sub-graphs of the first instance to
initiate the BSP, after the completion of whose supersteps, the \texttt{Compute} method is called on
all sub-graphs of the next instance for the next timestep iteration, and so on till the last graph
instance. So, while there is spatial concurrency across sub-graphs in a BSP superstep, each timestep
iteration is itself sequentially executed after the previous. In case of an \emph{independent}
pattern, the \texttt{Compute} method can be called on any graph instance independently, as long as
the BSP is run on each instance exactly once. The application terminates when all the BSP timesteps
complete. Here, we have both spatial concurrency across sub-graphs and temporal concurrency across
graph instances. An \emph{eventually dependent} pattern is similar, except that the
\texttt{Merge} method is called after the BSP timesteps complete on all instances of the graph collection. The
parallelism is similar to the independent pattern, except for the merge BSP which has to be
additionally executed last.

\textbf{User Logic.} The signatures of the \texttt{Compute} method and the
\texttt{Merge} method, in case of an \emph{eventually dependent} pattern, implemented by the user
are follows. The parameters are passed to these methods by the execution framework.
\begin{center}
\small
\texttt{\textbf{Compute}(SubgraphInstance \emph{sgInstance}, int \emph{timestep}, int \emph{superstep}, Message[] \emph{msgs})} \\
\texttt{\textbf{Merge}(SubgraphTemplate \emph{sgTemplate}, int \emph{superstep}, Message[] \emph{msgs})}
\end{center}

Here, the \texttt{SubgraphInstance} has the time variant attribute values of the corresponding graph
instance for this BSP, in addition to the sub-graph topology that is time invariant.  The
\texttt{\emph{timestep}} is a sequential number that corresponds to the graph instance's index,
while the \texttt{\emph{superstep}} corresponds to the superstep number inside the BSP execution. If
the superstep number is 1, it indicates the beginning of an instance's execution. Hence, it offers a
context for interpreting the list of messages, \texttt{\emph{msgs}}. In case of a \emph{sequentially
dependent} application pattern, messages received when \texttt{\emph{superstep=1}} have arrived from
its preceding BSP instance upon its completion. Hence, it indicates the completion of the previous
timestep, start of the next timestep and helps to pass the state from one instance to the next. If,
in addition, the \texttt{\emph{timestep=1}}, then this is the first BSP timestep and the messages
are the inputs passed to the application. For an \emph{independent} or \emph{eventually dependent}
pattern, messages received when the \texttt{\emph{superstep=1}} are application input messages since there
is no notion of a previous instance. In cases where \texttt{\emph{superstep>1}}, these are messages
received from the previous superstep inside a BSP.

\textbf{Message Passing.} Besides message passing between sub-graphs in supersteps, supported by the
sub-graph centric abstraction, the \texttt{Compute} and \texttt{Merge} methods can use these
additional message passing and application termination constructs, depending on the design pattern.
{\small \texttt{\textbf{SendToNextTimeStep}(Message \emph{msg})}}, used in sequentially
dependent pattern, passes message from a sub-graph to the next instance of the same sub-graph,
available at the start of the next timestep. This can be used to pass the end state of an instance
to the next instance.
{\small \texttt{\textbf{SendToSubgraphInNextTimeStep}(long \emph{sgid}, Message \emph{msg})}}, is similar,
but allows a message to be targeted to another sub-graph in the next timestep's instance. 
{\small \texttt{\textbf{SendMessageToMerge}(Message \emph{msg})}} is used in the eventually
dependent pattern by sub-graphs in any timestep to pass messages to the merge method, which will be
available after all timesteps complete execution.
{\small \texttt{\textbf{VoteToHalt}()}}, depending on context, can indicate the end of a BSP
timestep, or the end of the iBSP application in case this is the last timestep of a sequentially
dependent pattern. It is also used by the merge method to terminate the application.


\textbf{Gopher Framework.} Gopher is a distributed framework implementation of the sub-graph centric
BSP abstraction~\cite{simmhan:ipdps:2014}, which has been extended for the proposed iterative BSP
model. It supports the new user logic and messaging APIs mentioned above, and allows composition and
distributed execution of iBSP applications based on the three design patterns. Gopher works in
tandem with the GoFS distributed data storage for time-series graphs introduced next. This
cooperation utilizes some of the computational concurrency offered by the abstractions while also
leveraging the data locality present in GoFS.






\subsection{Sample iBSP Application}



Algorithm~\ref{alg:temporal-path} shows a sequentially dependent iBSP application that locates a
vehicle, based on its license place $\mathbb{V}$, within a road network and tracks the vehicle over
time across multiple instances. The graph template is a road network, and the graph instances have
vertex attributes with license plates of vehicles seen at the intersection, for the duration of that
instance (e.g. $5 mins$). The first timestep determines the vehicle's location in the entire graph
and traces it spatially across sub-graphs, using message passing across supersteps, it until it goes
missing in the instance's time duration. It then moves to the next timestep containing the instance
for the next $5 mins$, and resumes the traversal from the last known sub-graph location, using
message passing between timesteps. The algorithm terminates once all instances are exhausted.

\begin{algorithm}
\small
\caption{Temporal Path Traversal using iterative BSP}\label{alg:temporal-path}
\begin{algorithmic}[1]
\Procedure{Compute}{Instance sgi, int iteration, int superstep, Message msgs[]}
\State{searchRoots $\gets \varnothing $} \Comment{\emph{Vertices to begin search for $\mathbb{V}$ on}}
\If{superstep = 1} \Comment{\emph{Initialize from previous iteration}}
	\If{iteration = 1} \Comment{\emph{Initialize from user input}}
		\State v $\gets$ initial\_location
		\State searchRoots.add(v)
	\Else
        \Comment{\emph{Get the last vertex seen with the search value, $\mathbb{V}$, from the
                    previous iteration timestep.}}
		\State v $\gets (\operatorname*{arg\,max}_{m \in msgs} \lbrace \text{m.TimeStamp} \rbrace)$.vertex 
		\State searchRoots.add(v)
	\EndIf
\Else
\Comment{\emph{Process messages from previous superstep}}
	\ForAll{Message  m \textbf{in} msgs} 
		\State v $\gets$ m.vertex
		\State searchRoots.add(v)
	\EndFor	
\EndIf

\Comment{\emph{Depth first search on sub-graph from last seen location}}
\State $\langle$ remoteSet, foundLocs $\rangle \gets$ \Call{DFS}{sgi, searchDepth, $\mathbb{V}$}

\Comment{\emph{If DFS crosses to neighboring sub-graph, send message to remote sub-graph to continue search in next superstep}}
\ForAll{$\langle$ remoteSG, remoteVertex $\rangle$ \textbf{in} remoteSet }
   	\State Message m $\gets \langle$ remoteVertex $\rangle$ 
	\State{\Call{SendToSubGraph}{remoteSG, m}}
\EndFor

\If{Locations $\ne \varnothing$} \Comment{\emph{If current instance exhausted, continue search in next timestep}}
	\ForAll{$\langle$ vertex, TimeStamp $\rangle$ \textbf{in} remoteSet}
	\State \Comment{\emph{Send last known location message to next instance of the same sub-graph}}
  		\State Message m $\gets \langle$ vertex, TimeStamp $\rangle$ 
		\State{\Call{SendToNextTimeStep}{vertex, TimeStamp}}
	\EndFor
\EndIf

\State{\Call{VoteToHalt}{ }}

\EndProcedure

\Statex

\end{algorithmic}
\end{algorithm}

\section{Distributed Storage and Execution Patterns}
\label{sec:gofs}

Big data
applications can quickly become I/O bound unless the data storage and layout on disk are well
planned for the intended usage patterns. While advanced database schema planning has given way to flat, schema-free
distributed storage using HDFS and no-SQL databases, the interconnected nature of graph
datasets, with the additional temporal dimension considered here, pose challenges to tuple-based storage models.

We propose a Graph oriented File System (GoFS) for distributed storage of time-series graphs on
commodity clusters or Cloud VMs, with spinning disks. GoFS is architected for the data access patterns
associated with time-series graph analytics, though in effect, both the programming abstractions
and the data layout were co-designed. The typical usage model of GoFS is by Gopher, which loads
subgraphs in the local host's graph partition, and scans through instances as part of the BSP
timestep iterations.  The key tenets observed in this co-design are to: maximize concurrent execution,
minimize network data movement, reduce disk I/O, and increase the compute to I/O ratio. Given the
\emph{write once/read many} model of GoFS, we trade off data layout cost against improved runtime
performance. These choices are reflected in the GoFS data layout design and Gopher runtime execution
optimizations.

\begin{figure}
\centering
\includegraphics[width=\columnwidth]{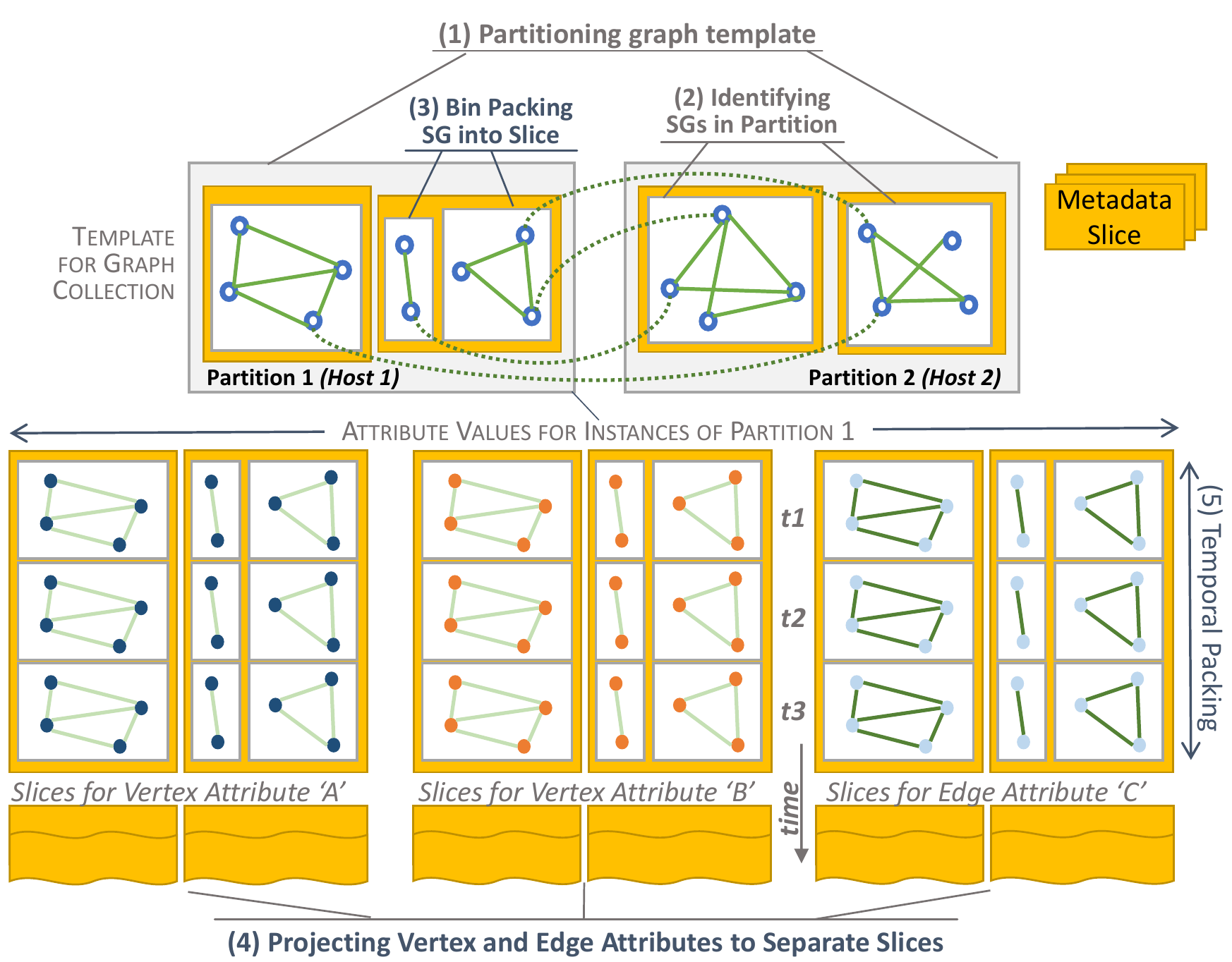}
\caption{GoFS Data Layout. Top shows graph template partitioned and mapped to slices on two
  hosts. Bottom shows time-series graph instances on Host 1. Yellow boxes are slices.}
\label{fig:gofs}
\vspace{-0.2in}
\end{figure}

\subsection{Partitioned Storage using Slices}
GoFS can store multiple time-series graph collections, though we limit our discussions to storing a
single collection for simplicity. A subgraph within the entire graph forms the unit of computation
for the BSP model, so maximizing the concurrent execution of subgraphs in an
instance is key. Since instances share the same topology, GoFS partitions the graph template into as
many partitions as the number of hosts in the cluster, and identifies one or more subgraphs for
a partition. So each host works on at least one subgraph for each instance.

Our default partitioning balances the number of vertices per partition and minimizes the remote edge
cuts. While this can translate to more even computational load per host and less messaging across
them, reality differs. Given that the unit of computation is a subgraph and hosts have multiple
cores, an additional partitioning goal should ensure equal number of uniform sizes subgraph per
partition, preferring the number of subgraphs as a multiple of the cores per host. This keeps all
cores busy with work that has similar time complexity. Intuitively, the BSP model is limited by the
slowest subgraph in each superstep which leads to idle workers~\cite{simmhan:ipdps:2014}. This
secondary goal is part of future work.

\emph{Slices} are the unit of storage and access from disk (Fig.~\ref{fig:gofs}). A slice translates to a single file
with a serialized graph data structure. Given that a single slice may contain many chunks of information, bulk reading of a slice at a time ensures that the disk
latency is amortized across a chunk of logically related bytes rather than performing random
access. Slice types vary, and may contain graph topology, attributes, metadata, and so on, as
discussed below. Given this diversity and the inherent irregularity of graph data, we allow slices
sizes to span a range ($O(MB)$) while retaining the disk latency:bandwidth benefits.

\subsection{Iteration, Filtering and Projection}
The large size of a time-series graph collection implies that we cannot retain it entirely in
distributed memory. At the same time, the time-series nature offers an inherent order to the
instances that applications may leverage (e.g. using a sequentially dependent pattern). As with
other tuple-based No-SQL storage, the GoFS access API provides an \emph{iterative model} to retrieve
graph instances in time order.  The API itself is subgraph centric; it provides an iterator over subgraphs
within the partition (space), and an iterator over instances for each subgraph (time).  
\begin{center}
\small
\texttt{Iterator<SubgraphTemplate> \textbf{Partition.getSubgraphs}()}\\
\texttt{Iterator<SubgraphInstance> \textbf{SubgraphTemplate.getInstances}(Time \emph{start}, Time \emph{end}, AttrName[]
  \emph{vertexAttrs}, AttrName[] \emph{edgeAttrs})}
\end{center}

\emph{Template slices}
capture the topology and attribute schema for the subgraphs in the partition. If the analytics is
scoped to just the graph structure, only the template slices need to be accessed.  The API
only operates on slices present on the local host and partition. This eliminates network transfer
at the GoFS layer at runtime and pushes cross-machine coordination to the Gopher application. So, within a
partition, a graph instance devolves to set of subgraph instances, described by their attribute
values. Given this API, a subgraph instance forms a logical unit of instance storage in the partition.

GoFS is a distributed graph \emph{data store}, not a graph database. While \emph{ad hoc} queries are
out of scope, the flexibility of a time-series graph model, with name-value pairs on
vertices and edges, means that applications are unlikely to access the entire collection on every
run. Two queries we support in the GoFS API are \emph{filtering} instances on time and \emph{projecting}
attributes. 

A collection may span many years or be stored at a very fine time granularity. While each instance can capture a
non-overlapping time \emph{duration} (rather than a strict moment in time), analytics will often
scan over a \emph{temporal subset} of instances. The GoFS access API allows a start and end
time to be passed in as arguments. A \emph{metadata slice} maintains an index from time ranges in the collection to
specific slices that contain data related to a range, limiting temporal queries to access only those
slices on disk.

Each vertex and edge may have many typed attributes, with corresponding values per
instance. Applications frequently need only a few of these attributes. The GoFS API lets
users pass vertex and edge attribute names whose values should be returned (projected) in the
subgraph instance. Rather
than store all attribute values for a subgraph instance on the same slice, we
maintain separate \emph{attribute slices} for each attribute of an instance, with a metadata slice
mapping the attribute name to the relevant slices (Fig.~\ref{fig:gofs}). This too helps localize disk access.

Lastly, we also support \emph{constant} and \emph{default} values to be specified for a vertex or an
edge attribute as part of its template schema. This allows non-changing or infrequently changing
attribute values to be stored just once in the template slice, and overridden (if a default value,
not if constant) by an instance. The GoFS API makes the value inheritance transparent. It also gives users the
ability to do more with just the graph template.







\subsection{Temporal Instance Packing}
Spatial partitioning into subgraphs and projecting of attributes into slices helps separate
independent units of concurrent executions by the design patterns while minimizing the disk
I/O. Now, we need to ensure aggregation of data within slices, based on their colocated execution
suggested by the design patterns. This allows us do more compute per slice I/O read from disk to
memory.

Co-temporal data is likely to show highly localized execution patterns. The \emph{iteration} in the BSP is
over graph instances over linear time. While a sequentially dependent pattern exhibits a causal
relationship over time with a time-ordered
execution, the other two patterns can also leverage (though they do not \emph{have} to) temporal
locality across independent BSP timesteps. Hence, instances that are temporally local will be
accessed in close proximity during execution. 

We take advantage of these patterns by \emph{packing} nearby instances together within a single
slice (Fig.~\ref{fig:gofs}). Thus, an attribute slice storing a subgraph instance values will contain adjacent instances,
and the slice will contain instances that span a time duration. So reading this slice from disk to
access one instance will effectively load a sequence of instances. If this slice is cached in memory (as we discuss
in \S~\ref{sec:caching}), operating over the next instance will not require a disk read. 

The number or time duration of instances packed into a slice can be tuned. But the key aspect is
that this value has to be consistent across all subgraph instances. The typical BSP application
loads and operates on all subgraphs of a graph instance. So if even one of them forces a slice read
due to skewed packing, the penalty will be paid by all.


\subsection{Subgraph Bin Packing and Ordered Iterators}
In an ideal partitioning, we would have exactly as many uniform-sized subgraphs per partition as the
number of cores in the host. In reality, partitioning large graphs results in partitions with
hundreds of subgraphs with highly variable vertex and edge counts (\S~\ref{sec:perf-data}). This
causes two problems: numerous slice files (sometimes, millions) and highly variable file sizes,
causing imbalances in slice read times across subgraphs and also imbalances in execution.
While we could pack more time ranges into slices or grouping multiple attributes into a
slice, the locality between these is weaker, thus wasting disk I/O or memory, and they fail to
address imbalances in computation between large and small subgraphs.

To ameliorate this problem we introducing a \emph{subgraph bin packing} scheme. Within a BSP timestep for a
graph instance, users will end up accessing all subgraphs in the partition. So there is high
topological locality. By having a fixed number of slices (bins) and packing multiple subgraphs into
a slice (bin) to balance the number of vertices/edges/vertices+edges in a bin, we limit the slice
size and count. While the GoFS API makes this binning opaque, it does suggest a balanced execution
order for a BSP by returning subgraphs in a \emph{bin-major order} through the partition iterator. 
This also ensures that spatial locality for slice access from disk is preserved, processing all
subgraphs in a bin before moving to the next.

%

\subsection{Slice Caching}
\label{sec:caching}
The net effect of these optimizations is that a single slice (file on disk) can contain information
for several subgraphs and for several instances, which are colocated based on expected access
patterns. To fully take advantage of this locality, GoFS caches slices in memory, once loaded from
disk, up to a predetermined number of slots. We use a least recently used algorithm for
cache eviction. The impact of this is a marked reduction in the number of
disk reads. The cache size is again configurable, and has to balance the memory needs of the
analytics application with the locality of its access instance pattern. The API makes the caching
transparent to the user. 

In summary, GoFS implements a number of data layout optimizations to leverage instance locality and
caching. The temporal packing and subgraph binning offer increased read efficiency and cache hits,
and decrease the number of files on disk and open handles. These optimizations are
targeted at graph instances, which are incrementally loaded based on the application's
design pattern. The graph template is loaded once and retained in memory, and hence has a fixed
overhead.






\section{Performance Analysis}
\label{sec:perf}

In this section, we empirically validate the use of the iBSP abstraction to construct analytics over
time-series graphs using Gopher. We also evaluate the impact of our proposed data layout optimizations in
GoFS, on various graph access patterns using micro-benchmarks, as well as on Gopher applications.




\subsection{Dataset and Application}
\label{sec:perf-data}

Time-series graphs are not yet collected and curated frequently in their natural form. Instead of
synthetically generating instance data from several widely available real graph
topologies~\footnote{Stanford Network Analysis Project, http://snap.stanford.edu}, we rather
use a single real world time-series graph dataset for our 
validation that captures temporal
snapshots of internet-work behavior. The graph template is a subset of the Internet constructed by
periodically sending network traceroutes from a dozen vantage hosts to 10~Million hosts around the
world. Destination hosts and intermediate routers form vertices in the template, identified by the
their IPv4 address, and edges represent hops in the trace. These traces are sent periodically to
measure the latency and bandwidth, among others, and a graph instance is created for every 2~hour
window. The vertices and edges have both static/slow changing attributes (e.g. IP address),
and fast changing ones (e.g. hop latency, destination IP), and zero or more values for each
during each 2~hour window, depending on the numbers of traces that passed through them during this period. We refer to this
traceroute-based time-series graph as \textbf{TR}.

In all, the template has $19,442,778$ vertices, $22,782,842$ edges, a diameter of $25$, and a
small-world structure reflecting the Internet topology. There are \emph{7~attributes each} for
vertices and edges, with boolean, integer, float and string types, and zero or more values per
attribute per vertex/edge. There are $146$ graph instances, each spanning a \emph{2~hour} window, and
covering a $12~day$ period of network statistics collection.

\begin{figure}
\centering
\includegraphics[width=\columnwidth]{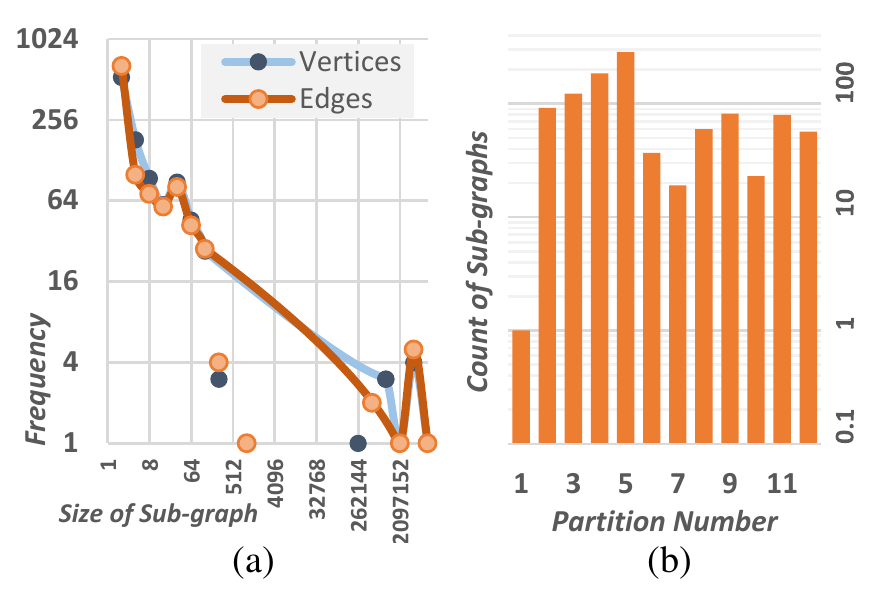}
\caption{Frequency distribution of (a) \# of vertices \& edges per sub-graph, and (b) \# of
  sub-graphs per partition. Log scale.}
\label{fig:data-distr}
\vspace{-0.2in}
\end{figure}

The TR time-series graph collection is partitioned across \emph{12 hosts} in a commodity
cluster. The partitions have between $1$ and $285$ sub-graph each, with the number of vertices/edge
per sub-graph ranging from $1/0$ to $5,803,661/6,511,896$ (Fig.~\ref{fig:data-distr}). As can be
imagined, there is an inverse correlation between the number of sub-graphs in a partition and the
sizes of sub-graphs in it.  Each host has an 8-core Intel Xeon CPU, 16~GB RAM and 1~TB SATA HDD and is inter-connected by a Gigabit Ethernet. The hosts run 64-bit Java 7 on Ubuntu Linux. 





We implement three different applications, that span the three design patterns emphasized in this paper: Single Source
Shortest Path (SSSP) \emph{(sequentially dependent)}, N-hop latency \emph{(eventually dependent)},
and PageRank (\emph{independent}). SSSP finds the shortest path from a source IP address for an
instance to all other IP addresses using the A$^\star$/Dijkstra's algorithm, with latency as the edge weight. These distances are
incrementally aggregated between instances. N-hop latency builds a histogram of latency times taken
to reach IPs that are 'N' hops from a source IP; we use $N=6$. These histograms are folded into a
composite in the merge step. Lastly, PageRank offers a form of network centrality, and is executed
on each instance independently by only considering edges that were active in a trace for that
instance's period.

These applications validate the ability to map meaningful time-series graph analytics to our design
patterns using the iBSP model. However, in the interests of space, we limit our detailed 
analysis to SSSP because: (1) it is a popular, well-understood algorithm that is
representative of other traversals; and (2) it uses the sequential design pattern
which is the most restrictive in terms of concurrency, hence outlying the system's behavior under
constraints. Furthermore, we can compare these results with prior ones for SSSP on a \emph{single
template graph} using a \emph{non-iterative} sub-graph centric BSP~\cite{simmhan:ipdps:2014}.

\subsection{Micro-benchmarks on GoFS Data Layout}
There are three aspects of GoFS layout optimizations that we investigate here: temporal packing of
instances, bin packing of sub-graphs within a partition, and caching. The configuration of the first
two have to be decided at deployment time since they impact slice creation, while the cache size can
be configured at runtime by the client using GoFS. We use 4 different deployments of the TR
time-series graph collection, which combines a temporal packing of either $1$ instance (\emph{i1})
or $20$ instances (\emph{i20}) per slice with a sub-graph bin packing with $20$ bins (\emph{s20}) or
$40$ bins (\emph{s40}) per partition. Note that \emph{i1} refers to no temporal instance packing,
while we do not consider the case without sub-graph bin packing since the performance degradation is
too high. We also run experiments with caching disabled (\emph{c0}), and caching enabled with $14$ slice
slots (\emph{c14}), where 14 slots are sufficient to fit at least one slice from each of the 14
attributes available for vertices and edges.

\begin{figure}
\centering
\includegraphics[width=\columnwidth]{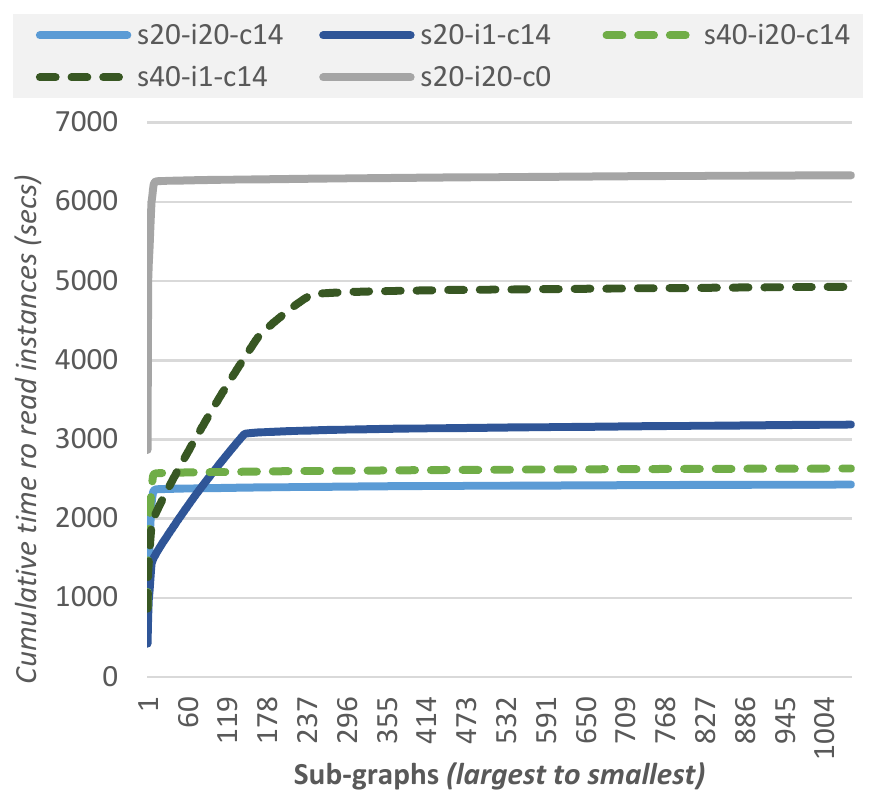}
\caption{Cumulative time of the total time to read all instances of each sub-graph. X Axis shows
  incremental sub-graphs being added, largest to smallest.}
\label{fig:cumu-time}
\vspace{-0.2in}
\end{figure}

As a high level comparison, for each of the deployments, we scan through all the sub-graphs, and for
each, we load all their instances. This translates to about 152,000 sub-graph instances read. We
then sum the total read time for all instances for each sub-graph, and plot this total read time
cumulatively for all the sub-graphs. Fig.~\ref{fig:cumu-time} shows this, with the sub-graphs on the
X axis sorted from largest to smallest. So $X=1$ identifies the time to read all 146 instances of
the largest sub-graph, while $X=1044$ is the total time to read all instances of all 1044 sub-graphs
across the 12 partitions. We plot the cached version for all deployments, and one non-cached
version is shown.

The plot illustrates the overall benefits of temporal packing, when we compare \emph{s20-i20-c14} and
\emph{s20-i1-c14} (light and dark blue solid lines), and \emph{s40-i20-c14} and \emph{s40-i1-c14}
(light and dark green dotted lines). For large sub-graphs on the left, the size of just a single
sub-graph instance on disk is large, and even without temporal packing (\emph{i=1}), the disk
latency cost is amortized over the loading time of the large, single-instance slice. With packing (\emph{i=20}),
we actually see poorer performance for large sub-graphs since the slice size is much larger for 20
instances and there may be file fragmentation and memory pressure effects. However, as we include
more modest sized sub-graphs, the benefits of temporal packing starts to be
exhibited. The cross over point is about 80 sub-graphs, for \emph{s20}, beyond which temporal
packing outperforms non-packing.

Similarly, for sub-graph bin packing, using 20 bins shows a marked benefit over 40 bins, with the benefits being larger when
temporal packing is not used. This is understandable since not using temporal packing and using a
large number of bins causes slice sizes to be smaller and the disk latency to dominate. With
temporal packing, there is a tangible but smaller difference between bin sizes of 20 and 40.
As the bin size increases, and tends towards the number of sub-graphs in the partition, this
degenerates to non-bin packing approach.

The impact of caching is apparent in the single gray solid line at the top shown for
\emph{s20-i20-c0}. It is almost three time as large as the cached version for
\emph{s20-i20-c14}. Hence, the benefits of temporal packing and sub-graph binning are reaped only
when combined with caching, as otherwise, they fail to leverage the pre-fetching benefits of
locality end up I/O bound.


\subsection{Application Benchmark of SSSP}

We evaluate the iBSP implementation of temporal traversal of SSSP over multiple instances using a
sequentially dependent pattern. The iBSP SSSP is run on three different configurations of the TR
time-series graph, \emph{s20-i20-c0}, \emph{s20-i1-c14} and \emph{s20-i20-c14}, which are
respectively temporal packing of 20 without caching, no temporal packing with caching, and temporal
packing and caching enabled, all with sub-graph binning enabled with 20 bins.

\begin{figure}
\centering
\includegraphics[width=0.8\columnwidth]{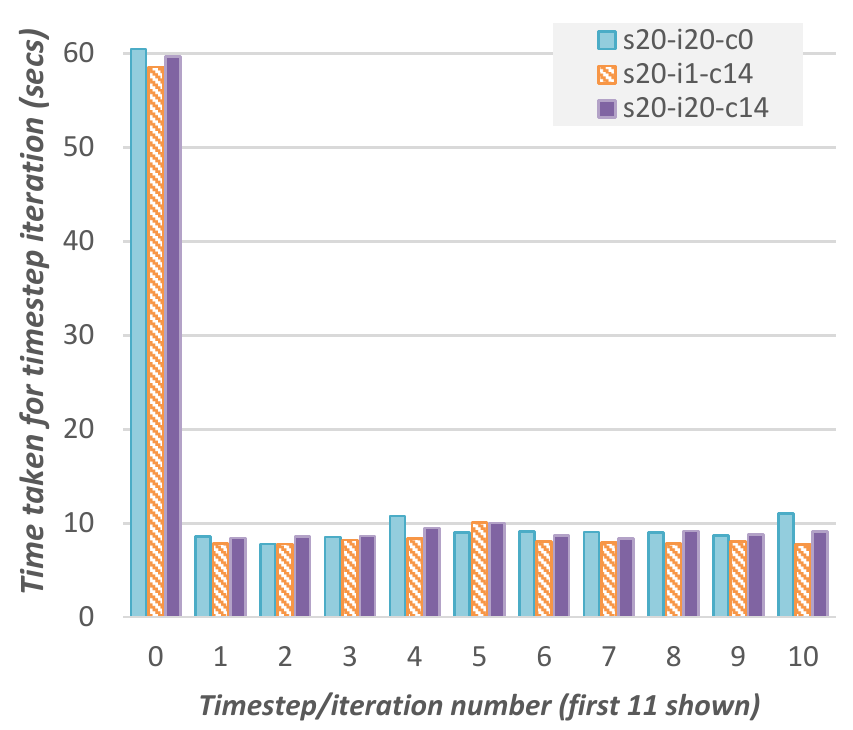}
\caption{Time per iteration (secs) for first 11 iterations of the iBSP SSSP application are shown for two
  deployments and cache sizes. Timestep 0 includes template load time.}
\label{fig:timestep}
\vspace{-0.2in}
\end{figure}

Fig.~\ref{fig:timestep} studies the time taken per timestep iteration, each corresponding to an SSSP
on one graph instance. The Y axis shows the total time taken by one BSP while the X axis show
sequentially increasing instances, with the first 11 being shown for conciseness. The bars in each
cluster refer to a different configuration of GoFS. We can see that the first timestep dominates,
and this is because the graph template is loaded as part of this timestep. Note that the template is
loaded just once at the start of the iBSP application. As we progress along the second and
subsequent timesteps, we see modest differences in the timings for each timestep for different GoFS
configurations. The penalty for not caching is evident in the first bar for \emph{s20-i20-c0}, while
the distinction between enabling temporal packing or not, in the second and third bars, is less
obvious. The domination of compute time for SSSP hides the differences, which also means that the
application is more compute bound than I/O bound, as we prefer.

\begin{figure}
\centering
\includegraphics[width=0.8\columnwidth]{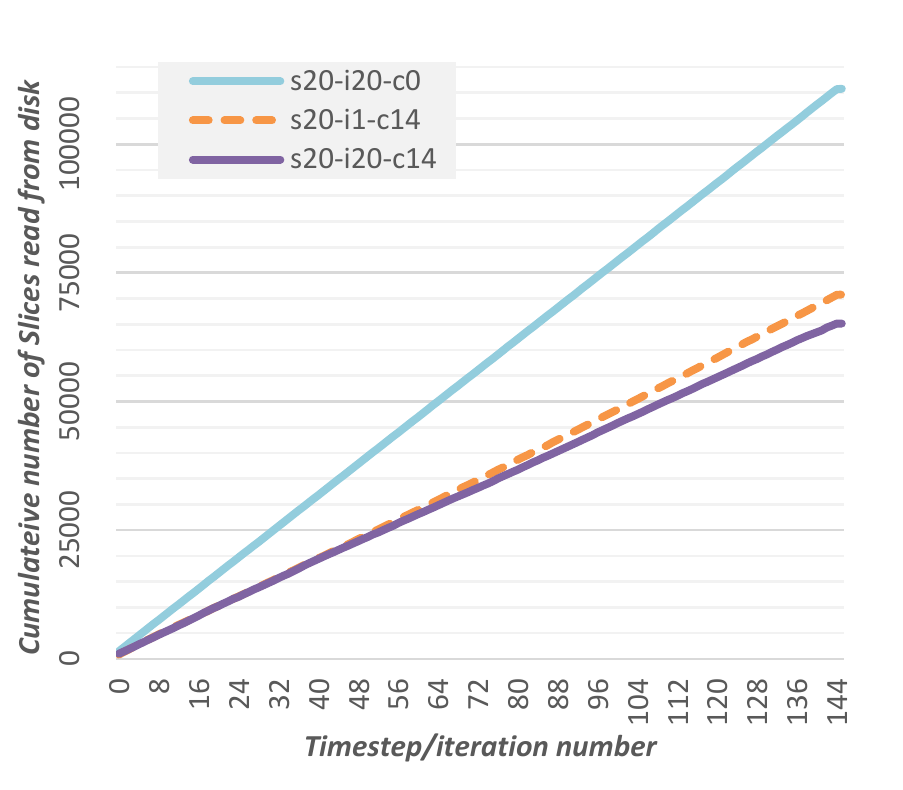}
\caption{Cumulative number of slices loaded (Y axis) for different deployments and caching, as the timesteps
  progress along X Axis for the iBSP SSSP application.}
\label{fig:slices}
\vspace{-0.2in}
\end{figure}

Fig.~\ref{fig:slices} offers a different view for this same experiment, from the perspective of the
cumulative number of slices that are read from disk as the timesteps progress. Here, the lack of
caching shows the high slope for the solid light blue line for \emph{s20-i20-c0}, while we see a
tangible difference in the number of slices read with and without temporal packing.

\section{Conclusions}
In summary, we have introduced and formalized the notion of time-series graph models as a first
class Big Data constituent that is of growing importance. We propose several design patterns for
composing analytics on top of this data model, and define an iterative BSP abstraction to define
such patterns for distributed execution. This leverages our existing work on sub-graph centric
programming for single graphs, and offers a high degree of parallelism, in space and in time. This
concurrency is made use of by Gopher which executes iBSP on commodity clusters, on conjunction with
the GoFS distributed graph data store, which is optimized for time-series graphs and the proposed
design patterns. The benefits of these are empirically validated for several GoFS configurations for
a canonical iterative SSSP application. These form a compelling basis for further investigation into
this novel Big Data and distributed abstractions space, with additional optimization problems open
to leverage the degrees of parallelism that we have exposed.


\bibliographystyle{IEEEtran}

\bibliography{main}

\end{document}